%% file: OneLoopIS-v10.tex
\documentclass[preprintnumbers,aps,prd,notitlepage,showpacs,nofootinbib,superscriptaddress]{revtex4}
\usepackage{amsfonts}
\usepackage{amsmath}
\usepackage{hyperref}
\usepackage{amssymb}
\usepackage[english]{babel}
\usepackage{graphicx}
\usepackage{epsfig}
\usepackage{bm}
\usepackage{longtable}
\usepackage{verbatim}
\usepackage{longtable}
\usepackage{subfigure}
\usepackage{tikz}
\usepackage{tikz-feynman}
\usepackage{pgffor}
\usepackage{multirow}
\usepackage{braket}

\newcommand{\mathsym}[1]{{}}

\tikzfeynmanset{compat=1.0.0}
\usetikzlibrary{arrows,shapes}
\usetikzlibrary{trees}
\usetikzlibrary{matrix,arrows} 
\usetikzlibrary{positioning}
\usetikzlibrary{calc,through}
\usetikzlibrary{decorations.pathreplacing}  
\usetikzlibrary{decorations.pathmorphing}
\usetikzlibrary{decorations.markings}

\tikzset{
    vector/.style={decorate, decoration={snake}, draw},
provector/.style={decorate, decoration={snake,amplitude=2.5pt}, draw},
antivector/.style={decorate, decoration={snake,amplitude=-2.5pt}, draw},
    fermion/.style={draw=black, postaction={decorate},
        decoration={markings,mark=at position .55 with {\arrow[draw=black]{>}}}},
    fermionbar/.style={draw=black, postaction={decorate},
        decoration={markings,mark=at position .55 with {\arrow[draw=black]{<}}}},
    fermionnoarrow/.style={draw=black},
    gluon/.style={decorate, draw=black,
        decoration={coil,amplitude=4pt, segment length=5pt}},
    scalar/.style={dashed,draw=black, postaction={decorate},
        decoration={markings,mark=at position .55 with {\arrow[draw=black]{>}}}},
    scalarbar/.style={dashed,draw=black, postaction={decorate},
        decoration={markings,mark=at position .55 with {\arrow[draw=black]{<}}}},
    scalarnoarrow/.style={dashed,draw=black},
    electron/.style={draw=black, postaction={decorate},
        decoration={markings,mark=at position .55 with {\arrow[draw=black]{>}}}},
bigvector/.style={decorate, decoration={snake,amplitude=4pt}, draw},
    line/.style={draw=black},
}\usetikzlibrary{decorations.markings}
\input{tcilatex}
\input{diagram}

\begin{document}

\title{\large{Dark Matter from a Radiative Inverse Seesaw Majoron Model}}

\author{Cesar Bonilla}
\affiliation{Departamento de F\'isica, Universidad Cat\'olica del Norte, Avenida
Angamos 0610, Casilla 1280, Antofagasta, Chile}
\email{cesar.bonilla@ucn.cl}
\author{A. E. C\'arcamo Hern\'andez}
\email{antonio.carcamo@usm.cl}
\affiliation{Universidad T\'ecnica Federico Santa Mar\'{\i}a, Casilla 110-V, Valpara\'{\i}%
so, Chile}
\affiliation{Centro Cient\'{\i}fico-Tecnol\'ogico de Valpara\'{\i}so, Casilla 110-V,
Valpara\'{\i}so, Chile}
\affiliation{Millennium Institute for Subatomic Physics at High-Energy Frontier (SAPHIR),
Fern\'andez Concha 700, Santiago, Chile}
\author{Basti\'an D\'{\i}az S\'aez}
\affiliation{Departamento de F\'isica, Universidad de Santiago de Chile\\
Casilla 307, Santiago, Chile}
\email{bastian.diaz.s@usach.cl}
\author{Sergey Kovalenko}
\affiliation{Departamento de Ciencias F\'isicas, Universidad Andr\'es Bello, Sazi\'e
2212, Piso 7, Santiago, Chile}
\affiliation{Centro Cient\'{\i}fico-Tecnol\'ogico de Valpara\'{\i}so, Casilla 110-V,
Valpara\'{\i}so, Chile}
\affiliation{Millennium Institute for Subatomic Physics at High-Energy Frontier (SAPHIR),
Fern\'andez Concha 700, Santiago, Chile}
\email{sergey.kovalenko@unab.cl}
\author{Juan Marchant Gonz\'{a}lez}
\email{juan.marchant@upla.cl}
\affiliation{Departamento de F\'{\i}sica, Universidad T\'ecnica Federico Santa Mar\'{\i}a%
\\
Casilla 110-V, Valpara\'{i}so, Chile}
\affiliation{Departamento de Matem\'atica, F\'isica y Computaci\'on, Facultad de Ciencias Naturales y Exactas, Universidad de Playa Ancha, Subida Leopoldo Carvallo 270, Valpara\'iso, Chile.}
\affiliation{Millennium Institute for Subatomic Physics at High-Energy Frontier (SAPHIR),
Fern\'andez Concha 700, Santiago, Chile}

\begin{abstract}
We propose a Majoron-like extension of the Standard Model with an extra global $U(1)_X$-symmetry where neutrino masses are generated through an inverse seesaw mechanism at the 1-loop level. In contrast to the tree-level inverse seesaw, our framework contains dark matter (DM) candidates stabilized by a residual $\mathcal{Z}_2$-symmetry surviving 
spontaneous breaking of the $U(1)_X$-group. 
We explore the case in which the DM is a Majorana fermion. Furthermore, we provide  parameter space regions allowed by current experimental constraints coming from the dark matter relic abundance, (in)direct detection, and charged lepton flavor violation.

\end{abstract}

\maketitle

\section{INTRODUCTION}
It took half a century to experimentally confirm with stunning accuracy every single part of what constitutes a unified description of the strong and electroweak interactions
dubbed the Standard Model (SM). Despite the revolution and success in particle physics, the SM is far from being the final description of our universe, leaving  many of its fundamental properties unexplained. For instance, 
it does not account for the origin of neutrino masses, lacks any hint of about $85\%$ of the "dark" matter budget of the universe and what
is the nature of this hidden sector, and does not explain the baryon asymmetry of 
the universe.
These, among other issues, lead us 
to think that the SM is at 
best a low-energy effective field theory
that belongs to a bigger framework.

The simplest and most popular realization to explain the smallness of neutrino masses 
is through the Type-I seesaw mechanism~\cite{Minkowski:1977sc,Yanagida:1979as,Glashow:1979nm,Mohapatra:1979ia,Gell-Mann:1979vob, Schechter:1980gr, Schechter:1981cv}. In this approach, Majorana right-handed neutrinos (RHN), $N_R$, are added to the SM. This implies to extend the Yukawa Lagrangian by including terms $y_\nu \bar{L}N_R \tilde{H} + M_R N_R\overline{N_{R}^c}$. If we consider order one Yukawa interactions, i.e. $y_\nu\sim \mathcal{O}(1)$, the size of neutrino masses get determined by RHN masses according to $m_\nu\approx v_{\Phi}^2/M_R$, where $v_\Phi$ is the Higgs vacuum expectation value (vev). Then, the new physics energy scale is expected to be at fifteen orders of magnitude above the electroweak one for $m_\nu\sim 0.1$~eV. This feature makes the Type-I seesaw mechanism inaccessible to current experimental sensitivities.

One of the main motivations to explore other scenarios accounting for neutrino masses\footnote{For a recent review on neutrino mass mechanisms at different energy scales see for instance~\cite{Cai:2017jrq}.} is that some manifest new physics signatures at energies around the TeVs, i.e. at the reach of current or upcoming experimental searches~\cite{Boucenna:2014zba}. One example of this is the so-called inverse seesaw model which is characterized by predicting significant lepton flavor violating rates~\cite{Wyler:1982dd,Mohapatra:1986bd,Gonzalez-Garcia:1988okv,Abada:2014vea}. 
This model introduces two Majorana fermion pairs to the SM, $N_{R_i}$ and $S_{L_j}$ ($i,j=1,2,3$). These fields transform as singlets under the SM gauge group and carry a lepton number of +1. Then, after electroweak symmetry breaking the Lagrangian in the neutrino sector is given by
\begin{equation}
    \mathcal{L_\nu}= m_D \overline{\nu}_L N_R + M \overline{N}_R S_L + \mu S^T_L C^{-1} S_L + h.c.,
  \label{eq:lag1}  
\end{equation}
where $m_D$ and $M$ are Dirac $3\times3$ matrices, while $\mu$ is a Majorana $3\times3$ matrix breaking lepton number explicitly. The latter can have a dynamical origin~\cite{Mohapatra:1979ia}. As usual, $C$ denotes the charge conjugation matrix. The neutrino mass matrix in the $(\nu_L, N_R, S_L)$ basis   turns out to be
\begin{equation}
M_{\nu}=\left( 
\begin{array}{ccc}
0 & m_{D} & 0 \\ 
m_{D}^{T} & 0 & M \\ 
0 & M^{T} & \mu 
\end{array}%
\right) .  \label{MnuIS}
\end{equation}%
Taking the limit $\mu_{ij}<<\left(m_D\right)_{ij}<< M_{ij}$ ($i,j=1,2,3$) leads to a $3\times3$ matrix for light neutrinos given by
\begin{equation}
    m_\nu\simeq m_D \frac{1}{M}\mu\frac{1}{M^T}m_D^T.
    \label{mnuIS0}
\end{equation}

Note that the lightness of neutrinos could be associated only to the smallness of $\mu$, which is naturally protected from large radiative corrections by the $U(1)_L$-symmetry of lepton number conservation, restored in the limit $\mu\to 0$ in Eq.~(\ref{eq:lag1}).
Obviously, neutrinos become massless in this limit. 
Nevertheless lepton flavor violating processes are still
allowed ~\cite{Gonzalez-Garcia:1991brm}.
Non-zero but small $\mu$ can be generated via radiative corrections opening up the intriguing possibility for
relation of neutrino mass generation to the dark sector~\cite{Ma:2009gu,Bazzocchi:2010dt,Kajiyama:2012xg,Okada:2012np,Law:2012mj,Ahriche:2016acx,CarcamoHernandez:2017cwi,CarcamoHernandez:2018hst,Mandal:2019oth,Rojas:2019llr,CarcamoHernandez:2019lhv,Hernandez:2021uxx,CarcamoHernandez:2021iat,Abada:2021yot,Bonilla:2023wok}.
 
In this paper we propose a variant of the inverse seesaw model where the $\mu$ term is generated at the 1-loop level after the spontaneous breaking of a global $U(1)_X$ symmetry. This Abelian continuous symmetry breaks down to a 
discrete subgroup $\mathcal{Z}_2$  which is an exact low energy symmetry which stabilizes the Dark Matter (DM) particle  candidates of our model. Then, we explore the viability of having as DM the lightest Majorana fermion involved in generation of the $\mu$ term in Eq.~\eqref{mnuIS0} at 1-loop level. We analyze the case in which this thermal relic communicates to the SM mainly via Higgs portal and provide constraints and prospects for direct and indirect detection in DM searches. Let us point out that here, we focus only on the neutrino and DM sectors, leaving the possibility of including the quark sector for subsequent studies.\\

This paper is organized as follows. In section~\ref{model} we provide the details of the model such as the particle content,
charge assignments and symmetry breaking. In addition, we describe 
the scalar potential and mass spectrum. In section~\ref{DM} the phenomenology of the fermionic dark matter candidate is studied. The implications of the model in charged lepton flavor
violation are discussed in section~\ref{clfvandlepto}. We
state our conclusions in section~\ref{sec:conc}.

\section{The model}
\label{model} 
We consider a model that adds to the SM, two complex scalars $\sigma $ and $\eta $ and six Majorana fermions $\nu _{R_k}$, $N_{R_k}$ and $\Omega _{R_k}$ ($k=1,2$). All these new fields are $SU(2)$ gauge singlets and carry neutral electric charge. In addition, the existence of a global $U(1) _{X}$ symmetry is assumed. This symmetry breaks down to a $\mathcal{Z}_2$ symmetry when the singlet scalar gets a vacuum expectation value (vev) $\langle\sigma\rangle=v_{\sigma}$. Table~\ref{tab:scalar-fermions} shows the charge assignments of scalars and leptons under the $SU\left( 2\right) _{L}\otimes U\left( 1\right) _{Y}\otimes U\left( 1\right) _{X}$ symmetry. 

\begin{table}[h!]
\centering
\begin{tabular}{|c|ccc|ccccc|} 
\hline
                         & $\Phi$ & $\sigma$ & $\eta$ & $L_{L_i}$ & $l_{R_i}$ & $\nu _{R_k}$ & $N_{R_k}$ & $\Omega _{R_k}$  \\ 
\hline\hline
$SU\left( 2\right) _{L}$ & $2$    & $1$      & $1$    & $2$       & $1$       & $1$          & $1$       & $1$              \\
$U\left( 1\right)_{Y}$   & $1/2$  & $0$      & $0$    & $-1/2$    & $-1$      & $0$          & $0$       & $0$              \\
$U\left( 1\right) _{X}$  & $0$    & $-1$     & $1/2$  & $-1$      & $-1$      & $-1$         & $1$       & $-1/2$           \\
\hline
\end{tabular}
\caption{Charge assignments of scalar and lepton fields. Here $i=1,2,3$ and $k=1,2$.}
\label{tab:scalar-fermions}
\end{table}

In fact, after electroweak symmetry breaking the unbroken symmetry is $SU(3)_C\otimes U(1)_{EM}\otimes \mathcal{Z}_2$ where $\mathcal{Z}_2$ turns out to be the symmetry that stabilizes the dark matter candidate of the theory. Schematically, the symmetry breaking chain goes as follows,
\begin{eqnarray}
&&\mathcal{G}=SU(3)_{C}\otimes SU\left( 2\right) _{L}\otimes U\left( 1\right) _{Y}\otimes U\left( 1\right) _{X}  \notag \\
&&\hspace{35mm}\Downarrow v_{\sigma }  \notag \\[0.12in]
&&\hspace{15mm}SU(3)_{C}\otimes SU\left( 2\right) _{L}\otimes U\left( 1\right)
_{Y}\times \mathcal{Z}_2  \notag \\[0.12in]
&&\hspace{35mm}\Downarrow v_{\Phi}  \notag \\[0.12in]
&&\hspace{15mm}SU(3)_{C}\otimes U\left( 1\right) _{Q}\otimes \mathcal{Z}_2
\end{eqnarray}

where the Higgs vev is represented by $\langle\Phi^0\rangle=v_\Phi$.\\

Given the charge assignments shown in Table~\ref{tab:scalar-fermions} we have that the singlet's vev is invariant under the following transformation, $e^{2\pi i \hat{X}}\langle\sigma\rangle=\langle\sigma\rangle$, where $\hat{X}$ is the $U(1)_X$ charge operator. 
This implies the existence of a residual discrete symmetry 
$(-1)^{2\hat{X}}\in\mathcal{Z}_2$
surviving spontaneous breaking of the global $U(1)_X$ group. 
Therefore, to all fields  
are assigned the corresponding $\mathcal{Z}_2$-parities 
$(-1)^{2\hat{Q}_X}$ according to their $U(1)_X$ charges $Q_X$ in Table~\ref{tab:scalar-fermions}.
The particles $\eta$ and $\Omega_{R_k}$ $(k=1,2)$  have odd $\mathcal{Z}_2$-parities and form the dark sector of the model.

\subsection*{Scalar sector}
\label{scalarsector}

The scalar potential invariant under the symmetry group $\mathcal{G}$ is given by

\begin{eqnarray}
\label{eq:pot}
V(\Phi ,\sigma ,\eta )&=& -\frac{\mu _{\Phi }^{2}}{2}|\Phi |^{2}+\frac{\lambda _{\Phi }}{2}%
|\Phi |^{4}-\frac{\mu _{\sigma }^{2}}{2}|\sigma |^{2}+\frac{\lambda _{\sigma
}}{2}|\sigma |^{4}+\frac{\mu _{\eta }^{2}}{2}|\eta |^{2}+\frac{\lambda
_{\eta }}{2}|\eta |^{4} \notag\\
&+&\lambda _{1}|\Phi |^{2}|\sigma |^{2}+\lambda _{2}|\Phi |^{2}|\eta
|^{2}+\lambda _{3}|\sigma |^{2}|\eta |^{2}+ \frac{\mu _{4}}{\sqrt{2}}\sigma \eta ^{2}+h.c., 
\end{eqnarray}%
where the quartic couplings $\lambda_a$ are dimensionless parameters whereas the $\mu_a$ are dimensionful. In our analysis we will impose perturbativity ($\lambda_a<\sqrt{4\pi}$) and the boundedness conditions given in Appendix~\ref{apx:stabilityconds}. \\

The singlet $\sigma$ and the neutral component of the doublet $\Phi=(\phi^+,\phi^0)^T$ acquire vacuum expectation
values (vevs). Here the singlet's vev $v_\sigma$ is responsible for the breaking of the global $U(1)_X$ symmetry while the double's vev $v_\Phi$ triggers electroweak symmetry breaking. Therefore we shift the fields as
\begin{equation}
\phi^0 =\frac{1}{\sqrt{2}}(v_{\Phi }+\phi_R+i\phi_I),\quad \sigma =\left( \frac{%
v_{\sigma }+\sigma_R+i\sigma_I }{\sqrt{2}}\right) .
\end{equation}%

Evaluating the second derivatives of the scalar potential at the minimum one finds the CP-even, $M_R^2$, and CP-odd, $M_I^2$, mass matrices. The CP-even mass matrix $M_R^2$ mixes $\phi_R$ and $\sigma_R$ and its eigenvalues correspond to the squared masses of the physical scalar states. They are
given by
\begin{equation}
\label{higgsmasses}
m_{h_{1},h_{2}}^{2}=\frac{1}{2}\left( \lambda _{\sigma }v_{\sigma
}^{2}+\lambda _{\Phi }v_{\Phi }^{2}\mp \frac{\lambda _{\sigma }v_{\sigma
}^{2}-\lambda _{\Phi }v_{\Phi }^{2}}{\cos 2\theta }\right) ,
\end{equation}%
where we identify $h_{1}$ with the 125 GeV Higgs boson, $v_{\Phi }=246$ GeV,
and the mixing angle $\theta $ fulfilling 
\begin{equation}
\label{higgsmixing}
\tan 2\theta =\frac{2\lambda _{1}v_{\Phi }v_{\sigma }}{\lambda _{\sigma
}v_{\sigma }^{2}-\lambda _{\Phi }v_{\Phi }^{2}}.
\end{equation}%
Moreover, the flavor and physical bases are connected through out the 
following relations,
\begin{eqnarray}
 \sigma_R&=&-h_{1}\sin \theta +h_{2}\cos \theta, \notag \\
 \phi_R&=& h_{1}\cos \theta +h_{2}\sin \theta. \label{eq:srot} 
\end{eqnarray}
The CP-odd mass matrix $M_I^2$ has two null eigenvalues. One of them corresponds to the would-be 
Goldstone boson which becomes the longitudinal component of the $Z$-boson by virtue of the Higgs 
mechanism. The other one is the physical Goldstone boson resulting from spontaneous breaking of the global $U(1)_X$ symmetry, similar to the singlet Majoron model in Ref.~\cite{Mohapatra:1979ia}. Notice that given that this Goldstone is an SM singlet, it is invisible, and its phenomenological impact in the model is not dangerous. It includes any cosmological manifestations. Let us also stress that the mass of this field is non-vanishing, at least due to the well-known quantum gravity effects. It can also gain mass due to non-perturbative QCD effects in an extended version of our model in which the fields in the quark sector have non-trivial $U(1)_X$ assignments entailing a mixed $(SU(3)_C)^2U(1)_X$-anomaly. This is what defines our Majoron model variant. 
The masses of $Z_2$-odd scalar components $\eta =\eta _{R}+i\eta_{I}$ turn out to be 
\begin{eqnarray}
m_{\eta _{R}}^{2} &=&\mu _{\eta }^{2}+\frac{1}{2}\left( \lambda _{2}v_{\Phi
}^{2}+\lambda _{3}v_{\sigma }^{2}\right) +\mu _{4}v_{\sigma },  \label{rel1}
\\
m_{\eta _{I}}^{2} &=&\mu _{\eta }^{2}+\frac{1}{2}\left( \lambda _{2}v_{\Phi
}^{2}+\lambda _{3}v_{\sigma }^{2}\right) -\mu _{4}v_{\sigma },
\end{eqnarray}%
where the mass splitting of these two components can be recasted as $\mu _{4}=(m_{\eta _{R}}^{2}-m_{\eta_{I}}^{2})/(2v_{\sigma })$. Note that $\eta _{R}$ and $\eta _{I}$ are degenerate when $\mu _{4}\rightarrow 0$. Here, the lightest of these two components, $\eta_R$
and $\eta_I$, can be the stable dark matter.\\

\subsection*{Neutrino sector}

\begin{figure}[h!]
\begin{center}
\includegraphics[width=0.5\textwidth]{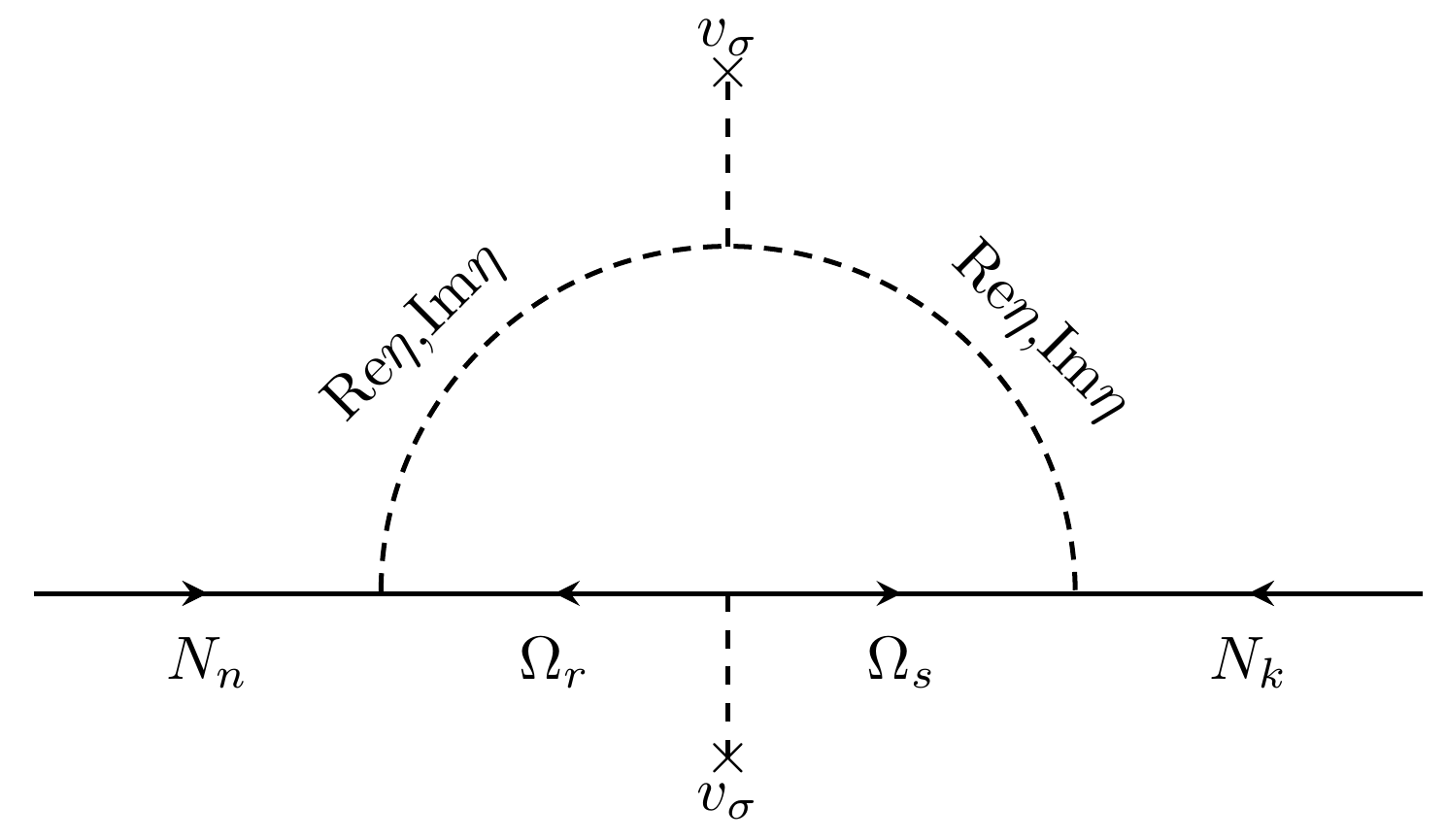}
\end{center}
\caption{One-loop Feynman diagram contributing to the Majorana neutrino mass
in Eq.~(\ref{Mnu}).}
\label{Neutrinoloopdiagram}
\end{figure}

Using Table~\ref{tab:scalar-fermions}, the invariant lepton Yukawa Lagrangian is given by
\begin{eqnarray}
\label{eq:yukL}
-\mathcal{L}_{Y}^{\left( l\right) } &=&\sum_{i=1}^{3}\sum_{j=1}^{3}\left(
y_{l}\right) _{ij}\overline{L}_{L_i}l_{R_j}\Phi+\sum_{i=1}^{3}\sum_{k=1}^{2}\left( y_{\nu }\right) _{ik}\overline{L}%
_{L_i}\nu _{R_k}\widetilde{\Phi }+\sum_{n=1}^{2}\sum_{k=1}^{2}M_{nk}\overline{%
\nu }_{R_n}N_{R_k}^{c}  \notag \\
&&+\sum_{n=1}^{2}\sum_{k=1}^{2}\left( y_{N}\right) _{nk}\overline{N}%
_{R_n} \Omega _{R_k}^{c}\eta+\sum_{n=1}^{2}\sum_{k=1}^{2}\left( y_{\Omega
}\right) _{nk}\overline{\Omega }_{R_n} \Omega _{R_k}^{c}\sigma+h.c.,  \notag
\end{eqnarray}

where $\psi^c=C \overline{\psi}^T$ and $\widetilde{\Phi }=i\sigma_2 \Phi^\ast$.
After spontaneous symmetry breaking (SSB), the neutrino mass matrix has the form,
 \begin{equation}
 M_{\nu }=\left( 
 \begin{array}{ccc}
 0_{3\times 3} & m_{D} & 0_{3\times 2} \\ 
 m_{D}^{T} & 0_{2\times 2} & M \\ 
 0_{2\times 3} & M^{T} & \mu 
 \end{array}%
 \right) ,  \label{Mnu}
 \end{equation}%
 where $m_D$ is the tree-level Dirac mass term 
 \begin{equation}
 \left( m_{D}\right) _{ik} =\left( y_{\nu }\right) _{ik}\frac{v_{\Phi }%
 }{\sqrt{2}},
 \label{ec:mdirac}
 \end{equation}
 with $i=1,2,3$ and $k=1,2$. The submatrix $\mu$ in Eq.~(\ref{Mnu}) is generated at
 one-loop level, 
 \begin{eqnarray} 
 \label{muloop}
 \mu _{sp} &=&\sum_{k=1}^{2}\frac{\left( y_{N}\right) _{sk}\left(
 y_{N}^{T}\right) _{kp}m_{\Omega _{k}}}{16\pi ^{2}}\left[ \frac{m_{\eta
 _{R}}^{2}}{m_{\eta _{R}}^{2}-m_{\Omega _{k}}^{2}}\ln \left( \frac{m_{\eta
 _{R}}^{2}}{m_{\Omega _{k}}^{2}}\right) -\frac{m_{\eta _{I}}^{2}}{m_{\eta
 _{I}}^{2}-m_{\Omega _{k}}^{2}}\ln \left( \frac{m_{\eta _{I}}^{2}}{m_{\Omega
 _{k}}^{2}}\right) \right] , 
 \label{eq:matrix-mu}
 \end{eqnarray}
with $s,p=1,2$. The Feynman diagram of $\mu$ is depicted in Figure~\ref{Neutrinoloopdiagram}.
One can see from Eq.~\eqref{muloop} that the $\mu$ term vanishes when the 
scalars $\eta _{R}$ and $\eta _{I}$ are degenerate. This implies that neutrino masses
go to zero in the limit $\mu\rightarrow 0$. 

Then one has that active light neutrino masses are generated via an inverse seesaw
mechanism at the one-loop level. Physical neutrino mass matrices are
given by\footnote{The diagonalization of the neutrino mass matrix in Eq.~(\ref{Mnu}) can be followed from Ref.~\cite{Catano:2012kw}}: 
\begin{eqnarray}
\label{Mnutilde}
\widetilde{M}_{\nu } &=&m_{D}\left( M^{T}\right) ^{-1}\mu M^{-1}m_{D}^{T},\hspace{0.7cm}   \\
M_{\nu }^{\left( -\right) } &=&-\frac{1}{2}\left( M+M^{T}\right) +\frac{1}{2}%
\mu ,\hspace{0.7cm} \label{Mnuminus} \\
M_{\nu }^{\left( +\right) } &=&\frac{1}{2}\left( M+M^{T}\right) +\frac{1}{2}%
\mu \label{Mnuplus}.  
\end{eqnarray}%
Now $\widetilde{M}_{\nu}$ is the mass matrix for active light
neutrinos ($\nu _{a}$), whereas $M_{\nu }^{(-)}$ and $M_{\nu }^{(+)}$ are
the mass matrices for sterile neutrinos. From Eq.~(\ref{Mnutilde}) one can see that 
active light neutrinos are massless in the limit $\mu\to0$ which implies that lepton number is a conserved quantity. Eqs.~(\ref{Mnuminus}) and (\ref{Mnuplus}) tell us that the smallness of
the parameter $\mu$ (small mass splitting) induces pseudo-Dirac pairs of sterile neutrinos.\\ 

\begin{figure}[h!]
\centering
\includegraphics[width=0.9\textwidth]{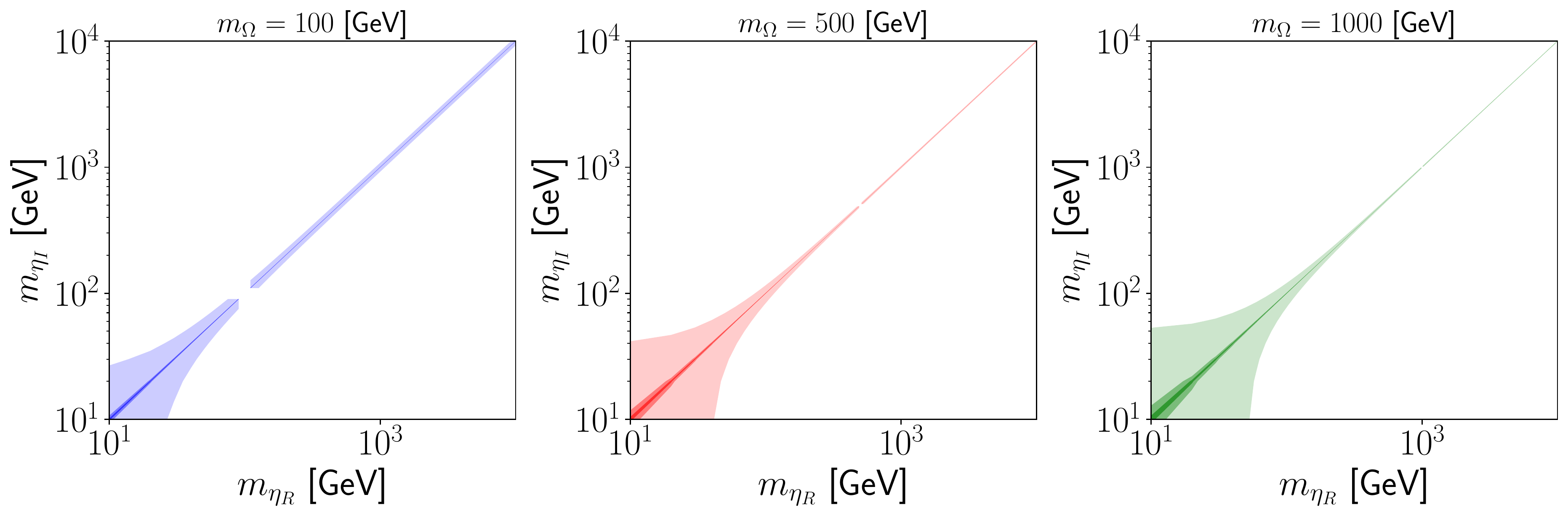}
\caption{Parameter space fulfilling $-10$ keV $\leq \mu \leq 10$ keV, for the DM masses indicated above each plot. The color in each plot, from light to dark, represents $y_N = 0.01, 0.05$ and 0.1, respectively. Here we
assume that $m_{\Omega_2} \gg m_\Omega$. The discontinuity appears when a degenerate mass spectrum is reached in Eq.~(\ref{muloop}).}
\label{fig:etaI-etaR}
\end{figure}

From Eq.~(\ref{Mnutilde}), one can see that a sub-eV neutrino mass scale can be linked to a small lepton number breaking parameter $\mu$ which depends on the Yukawas $y_N^2$, $y_\Omega$  and the masses of the particles running in the loop $(m_{\eta_R},m_{\eta_I},m_\Omega)$. This parameter is further suppressed by the loop factor, see Eq.~(\ref{muloop}). Figure~\ref{fig:etaI-etaR} shows the allowed parameter space regions for fixed Yukawa couplings $y_N$ and masses of the $Z_2$-odd Majorana fermions $\Omega _{k}$. Each plot is generated using Eq.~(\ref{muloop}), varying the masses $(m_{\eta_R},m_{\eta_I})$, fixing $m_\Omega$ and $y_N$. Then, from left to right, Figure~\ref{fig:etaI-etaR} shows the parameter space that fulfills $-10$ keV $\leq \mu \leq 10$ keV in the $(m_{\eta_R},m_{\eta_I})$-plane, considering $m_\Omega=100, 500,$ and  $1000$~GeV, respectively. In all panels, $y_N=0.01,0.05,$ and $0.1$, the smaller the Yukawa value, the lighter the region. The discontinuity appears when the mass spectrum in Eq.~(\ref{muloop}) is degenerate.

 As we have mentioned, the dark sector is formed by the $Z_2$-odd particles, see Table~\ref{tab:scalar-fermions}. The dark matter candidate of the model is the lightest component of either the singlet scalar $\eta$ or the Majorana fermion $\Omega$. 
The phenomenological consequences of having the lightest component of the scalar singlet $\eta$ as dark matter candidate is similar to what have been discussed in Refs.~\cite{Ma:2006km,Ahriche:2016acx,Bernal:2017xat,Mandal:2019oth,Abada:2021yot}. For this reason, in what follows we discuss only the constraints and projections of the model for the case in which the DM candidate is the Majorana fermion $\Omega$. 

\section{Fermion Dark Matter}
\label{DM}
\begin{figure}[t!]
\centering
\begin{tikzpicture}[line width=1.0 pt, scale=0.6]

\begin{scope}[shift={(0,0)}]
	\draw[fermion](-2.5,1) -- (-1,0);
	\draw[fermionbar](-2.5,-1) -- (-1,0);
	\draw[scalarnoarrow](-1,0) -- (1,0);
	\draw[line](1,0) -- (2.5,1);
	\draw[line](1,0) -- (2.5,-1);
    \node at (-3,1.0) {$\Omega$};
	\node at (-3,-1.0) {$\Omega$};
    \node at (-0.1,0.46) {$(h_1,h_2)$};
	\node at (1.4,1.3) {SM};
    \node at (1.4,-1.3) {SM}; 
    \node at (-0,-3) {$\textit{(a)}$};
\end{scope}

\begin{scope}[shift={(7,0)}]
	\draw[fermion](-2.5,1) -- (-1,0);
	\draw[fermionbar](-2.5,-1) -- (-1,0);
	\draw[scalarnoarrow](-1,0) -- (1,0);
	\draw[scalarnoarrow](1,0) -- (2.5,1);
	\draw[scalarnoarrow](1,0) -- (2.5,-1);
    \node at (-3,1.0) {$\Omega$};
	\node at (-3,-1.0) {$\Omega$};
    \node at (-0.1,0.46) {$(h_1,h_2)$};
	\node at (1.4,1.1) {$\chi$};
    \node at (1.4,-1.1) {$\chi$}; 
    \node at (-0,-3) {$\textit{(b)}$};
\end{scope}

\begin{scope}[shift={(14,0)}]
	\draw[fermion](-2.5,1) -- (-1,0);
	\draw[fermionbar](-2.5,-1) -- (-1,0);
	\draw[scalarnoarrow](-1,0) -- (1,0);
	\draw[scalarnoarrow](1,0) -- (2.5,1);
	\draw[scalarnoarrow](1,0) -- (2.5,-1);
    \node at (-3,1.0) {$\Omega$};
	\node at (-3,-1.0) {$\Omega$};
    \node at (-0.1,0.46) {$\chi$};
	\node at (1.4,1.1) {$\chi$};
    \node at (1.4,-1.1) {$(h_1,h_2)$}; 
    \node at (-0,-3) {$\textit{(c)}$};
\end{scope}

\begin{scope}[shift={(21,0)}]
 \draw[fermion](-3,1) -- (-1,1);
	\draw[fermionbar](-3,-1) -- (-1,-1);
	\draw[fermion](-1,1) -- (-1,-1);
	\draw[scalarnoarrow](-1,1) -- (1,1);
	\draw[scalarnoarrow](-1,-1) -- (1,-1);
    \node at (-3.6,1.0) {$\Omega$};
	\node at (-3.6,-1.0) {$\Omega$};
    \node at (-0.4,0) {$\Omega$};
	\node at (1.5,1) {$\sigma$};
    \node at (1.5,-1) {$\sigma$};   
    \node at (-0.8,-3) {$\textit{(d)}$};
\end{scope}

\begin{scope}[shift={(0,-6)}]
 \draw[fermion](-3,1) -- (-1,1);
	\draw[fermionbar](-3,-1) -- (-1,-1);
	\draw[scalarnoarrow](-1,1) -- (-1,-1);
	\draw[fermion](-1,1) -- (1,1);
	\draw[fermionbar](-1,-1) -- (1,-1);
    \node at (-3.6,1.0) {$\Omega$};
	\node at (-3.6,-1.0) {$\Omega$};
    \node at (-0,0) {$\eta_{R, I}$};
	\node at (1.8,1) {$N_R$};
    \node at (1.8,-1) {$N_R$};   
    \node at (-0.8,-3) {$\textit{(e)}$};
\end{scope}

\begin{scope}[shift={(7,-6)}]
 \draw[fermion](-3,1) -- (-1,1);
	\draw[fermionbar](-3,-1) -- (-1,-1);
	\draw[fermion](-1,1) -- (-1,-1);
	\draw[scalarnoarrow](-1,1) -- (1,1);
	\draw[scalarnoarrow](-1,-1) -- (1,-1);
    \node at (-3.6,1.0) {$\Omega$};
	\node at (-3.6,-1.0) {$\Omega$};
    \node at (-0.4,0) {$N_R$};
	\node at (1.8,1) {$\eta_{R,I}$};
    \node at (1.8,-1) {$\eta_{R,I}$};   
    \node at (-0.8,-3) {$\textit{(f)}$};
\end{scope}

\begin{scope}[shift={(14,-6)}]
 \draw[fermion](-3,1) -- (-1,1);
	\draw[scalarnoarrow](-3,-1) -- (-1,-1);
	\draw[scalarnoarrow](-1,1) -- (-1,-1);
	\draw[fermion](-1,1) -- (1,1);
	\draw[scalarnoarrow](-1,-1) -- (1,-1);
    \node at (-3.6,1.0) {$\Omega$};
	\node at (-3.6,-1.0) {$\eta_{R,I}$};
    \node at (-0.2,0) {$\eta_{R,I}$};
	\node at (1.8,1) {$N_R$};
    \node at (1.8,-1) {$\sigma$};   
    \node at (-0.8,-3) {$\textit{(g)}$};
\end{scope}

\begin{scope}[shift={(21,-6)}]
 \draw[fermion](-3,1) -- (-1,1);
	\draw[fermionbar](-3,-1) -- (-1,-1);
	\draw[scalarnoarrow](-1,1) -- (-1,-1);
	\draw[fermion](-1,1) -- (1,1);
	\draw[fermion](-1,-1) -- (1,-1);
    \node at (-3.6,1.0) {$\Omega$};
	\node at (-3.6,-1.0) {$N$};
    \node at (0.2,0) {$(h_1,h_2)$};
	\node at (1.6,1) {$\Omega$};
    \node at (1.6,-1) {$N$};   
    \node at (-0.8,-3) {$\textit{(h)}$};
\end{scope}

\end{tikzpicture}
\caption{\textit{Diagrams (a)-(g) are relevant for the freeze-out of $\Omega$. The diagram (h) is relevant for direct detection, with $N$ representing the nucleons. Here to simplify notation we have used $\sigma$ to denote either of $(h_1, h_2, \chi)$.}}\label{diagrams2}
\end{figure}
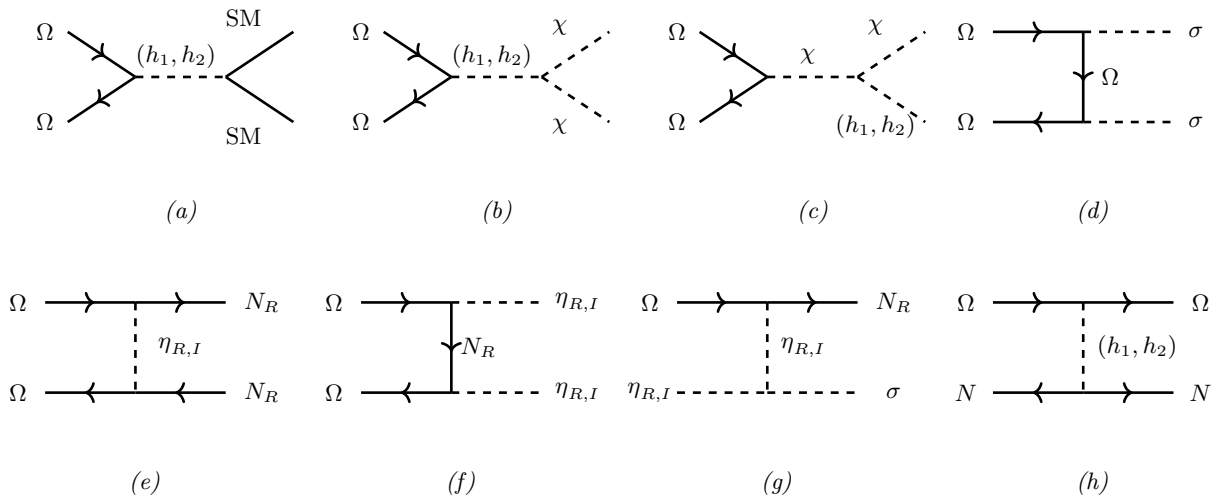


For simplicity, we consider the case in which $y_\Omega$ in Eq.~(\ref{eq:yukL}) is a diagonal matrix  and assume that $\Omega_{R_1}$ is the lightest $\mathcal{Z}_2$-odd state. That is, $\Omega_{R_1}$ is the fermion DM candidate accounting for the 80\% of the matter content of the universe. According to the Planck collaboration the DM relic abundance is~\cite{planck2018}\begin{equation}
\label{DMrelic}
\Omega_c h^2=0.1200\pm 0.0012\ \ \text{at} \ \ 68\% \text{C.L.}    
\end{equation}
In our setup  the Lagrangian providing the relevant interactions of $\Omega_{R_1}$ is given by
\begin{eqnarray}\label{omegalag}
 \mathcal{L} &\supset & y_{N_1}\bar{N}_R \Omega_{R_1}^c \eta+ y_{\Omega_1}\overline{\Omega}_{R_1} \Omega_{R_1}^c \sigma+ h.c.. 
 \end{eqnarray}

After SSB Eq.~(\ref{omegalag}) becomes
\begin{eqnarray}
 \mathcal{L} \supset (y_{N1}\bar{N}_{R} \Omega^c_{1R} \eta  + h.c.) + m_{\Omega 1}\overline{\Omega}\Omega + y_{\Omega 1}\overline{\Omega}\Omega(-h_1\sin\theta + h_2\cos\theta ) + y_{\Omega 1}i\overline{\Omega}\gamma^5\Omega \chi ,
\end{eqnarray}
where $m_{\Omega_1} = y_{\Omega_1} v_\sigma/\sqrt{2}$. We have defined $\Omega \equiv (\Omega_{1R})^c + \Omega_{1R}$, $N_R \equiv (-N_R^{+} + N_R^{-})/\sqrt{2}$ and $\chi\equiv\sigma_I$.\\

Taking into account the assumptions above, the relic abundance of $\Omega$ is determined by the annihilation channels shown in Figure~\ref{diagrams2}. Given that $\Omega$ is the DM candidate of the theory then $m_\Omega < (m_{N_R}, m_{\eta_R}, m_{\eta_I})$. In this case, the main annihilation channels are those s-channels depicted by diagrams (a), (b) and (c) in Figure~\ref{diagrams2}. We further simplify the analysis by considering that the annihilation channels mediated by the Higgs via the dimensionless parameters $\lambda_2$ and $\lambda_3$ are subleading. That is, we set $\lambda_2 = \lambda_3 = 0$. 
Therefore, the independent parameters to be used in the numerical analysis turn out to be $(m_\Omega, m_{\eta_R}, m_{\eta_I}, m_{h_2}, m_{N_R}, y_{N_1}, y_{\Omega_1}, \theta)$.\\

Let us note that in this inverse seesaw model, the Majorana dark matter candidate can interact with the atomic nucleons at tree-level. For this reason, our model gets restricted by direct detection constraints. As a matter of fact, these constraints come from the $t$-channel exchange of $h_1$ and $h_2$ shown by Figure~\ref{diagrams2}-(h). Then, here the spin-independent (SI) tree-level DM-nucleon scattering cross section is, approximately,~\cite{Garcia-Cely:2013wda, Garcia-Cely:2013nin} 
\begin{eqnarray}\label{dd}
 \sigma_\Omega \approx \frac{f_p^2 m_N^4m_\Omega^2}{4\pi v_\Phi^2(m_\Omega + m_N)^2}\left(\frac{1}{m_{h_1}^2} - \frac{1}{m_{h_2}^2}\right)^2(y_{\Omega_1} \sin 2\theta)^2,
\end{eqnarray}
where $m_N$ denotes the nucleon mass and the nuclear elements $f_p \approx 0.27$. The approximation given in Eq.~(\ref{dd}) does not take into account the finite width of both Higgs scalars, although the outputs of this expression match the numerical results from Micromegas code v5.3.35~\cite{Belanger:2013ywg} which do include these finite widths.

\subsection{Analysis and results}

In what follows, we compute the relic abundance of the Majorana fermion $\Omega$ assuming freeze-out mechanism, the direct detection via non-relativistic scattering, and indirect detection prospects today. For our calculation, we make use of Micromegas code v5.3.35~\cite{Belanger:2013ywg}.

\begin{figure}[t!]
\centering
\includegraphics[width=0.42\textwidth]{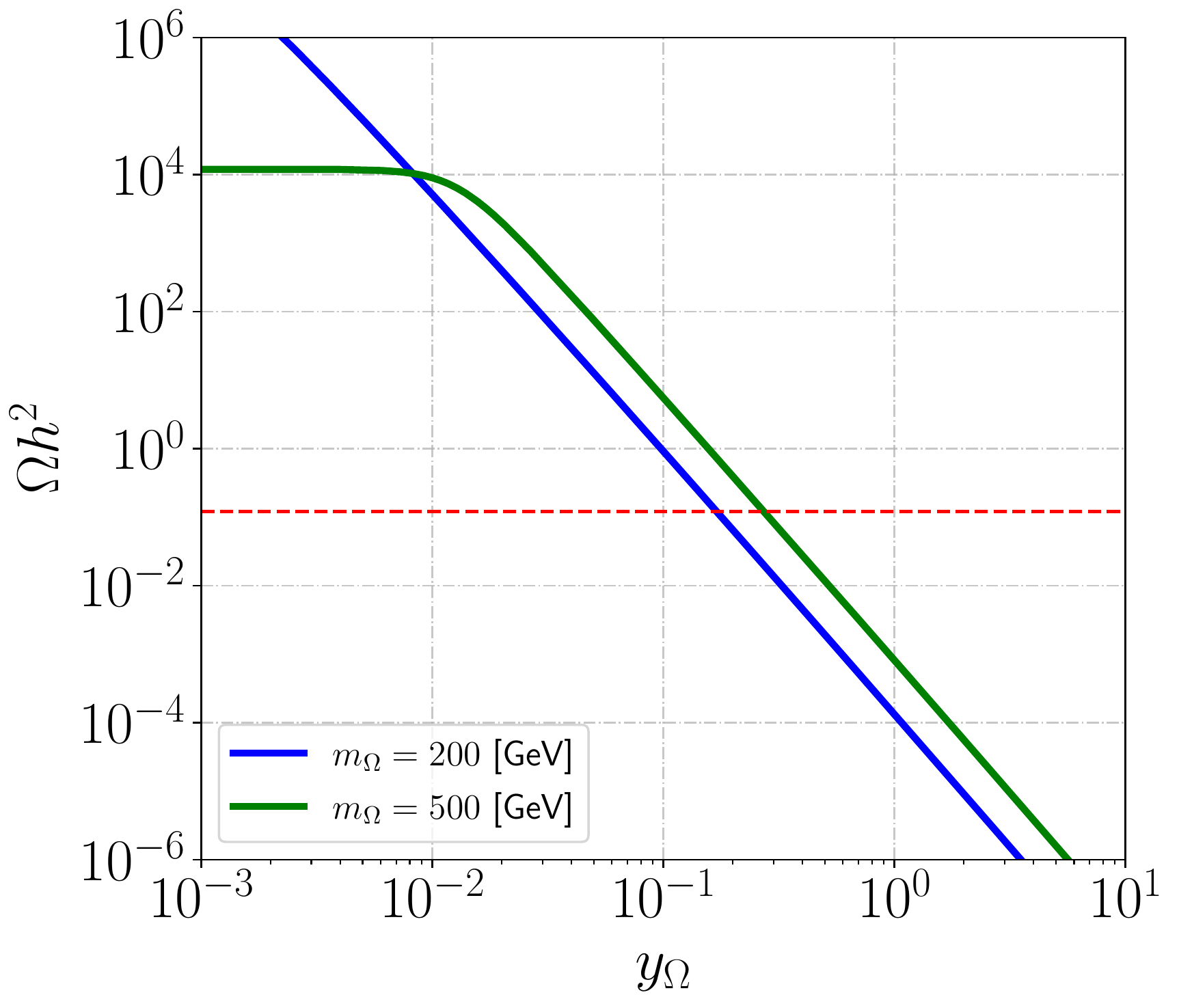}\quad
\includegraphics[width=0.42\textwidth]{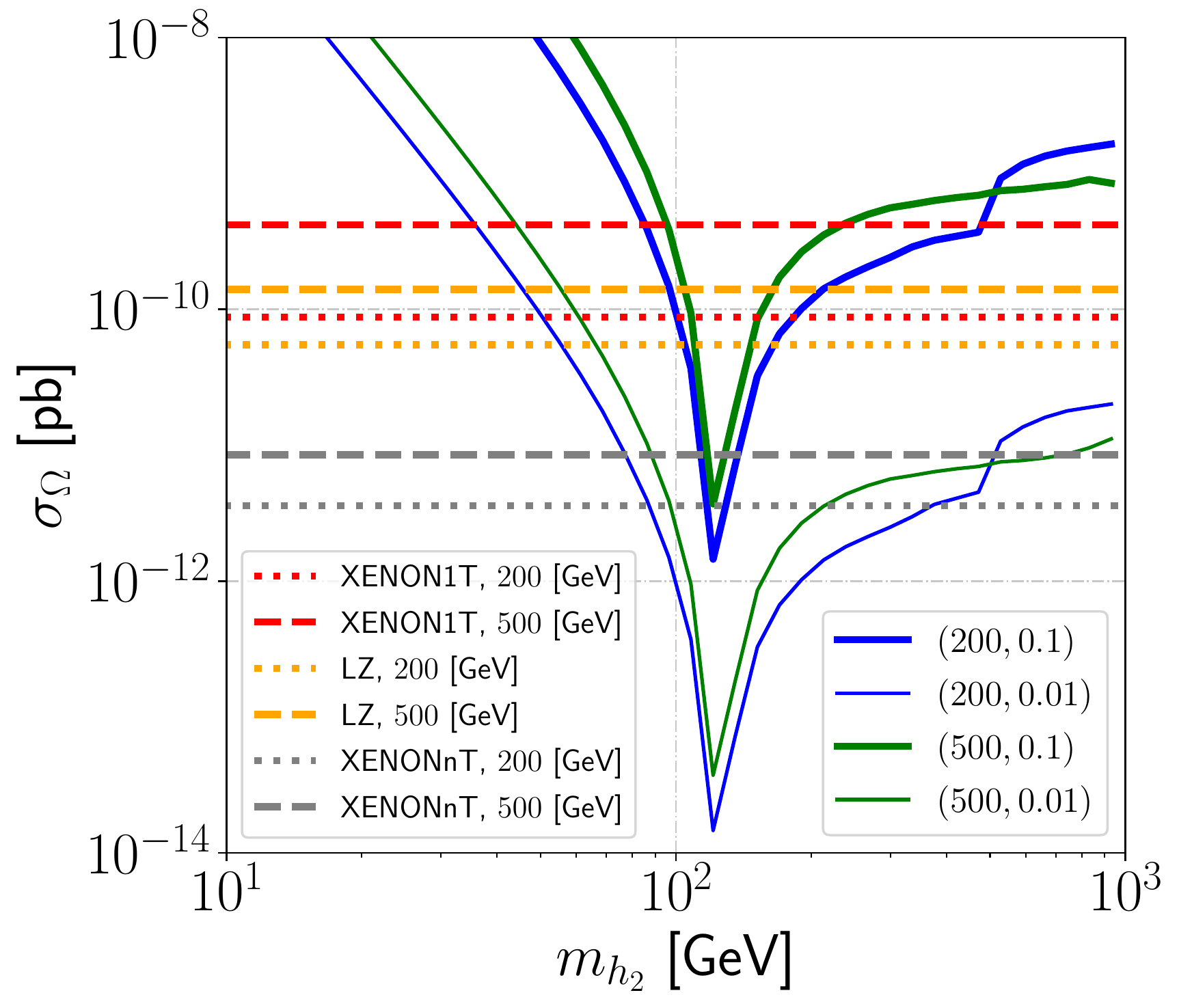}
\caption{Left-panel: Relic abundance as a function of $y_{\Omega}$ for $m_\Omega = 200$ and 500 GeV. Here we have set $m_{h_2} = 120$ GeV and $\theta = 0.1$, with the rest of the parameters specified in the text. Right-panel: Direct detection cross section as a function of $m_{h_2}$ and different combinations of $(m_\Omega \text{[GeV]}, \tan\theta)$. The horizontal red and orange lines are the current DD upper limits for each value of $m_\Omega$, whereas the grey curves are the XENONnT projections. The solid curves depicted here fulfill the correct relic abundance.}
\label{figb}
\end{figure}

As mentioned, here we explore the case where $\lambda_{1}\neq0$, $\lambda_2 = \lambda_3 = 0$. Therefore, the contribution to the relic abundance coming from the annihilation of $\Omega$ into SM particles happens only via the Higgs portal associated to $\lambda_1$. 
The left panel in Figure~\ref{figb} shows the relic abundance of $\Omega$ as a function of $y_\Omega$, for $m_\Omega = 200$ (solid blue) and $500$ GeV (solid greed), assuming $m_{h_2}= 120$ GeV and $\theta = 0.1$\footnote{Collider searches of additional scalars restrict the doublet-singlet mixing angle to be $\theta \lesssim 0.2$~\cite{Falkowski:2015iwa}}. This benchmark considers $y_N = 0.1$, $m_{\eta_R} = 2000$ GeV, $m_{\eta_I} = 2001$ GeV and $m_N = 300$ GeV (these parameters will be fixed to such values from now on). The limit of the DM relic abundance given in Eq.~(\ref{DMrelic}) is represented by the red dashed line. One can see from Figure~\ref{figb} that the relic abundance has a strong dependence on the DM mass and the parameter $y_\Omega$. For a small Yukawa coupling $y_\Omega \lesssim 10^{-2}$ and $m_\Omega > m_N$, the process $\Omega\Omega \rightarrow N N$ is kinematically allowed and dominates over the other annihilation channels. In the case of $m_\Omega < m_N$, the leading contributions to the relic abundance are those processes which involve the fields $(\chi, h_1, h_2)$ (diagrams $b$ and $c$ in Figure~\ref{diagrams2}). In such a case, the relic abundance turns out to be inversely proportional to $y_\Omega^2$. For this reason, as shown on the left-panel in Figure~\ref{figb}, the solid blue (green) curve decreases when the value of Yukawa increases. It is evident that, for a given $y_\Omega$, the relic abundance grows as the DM mass decreases. This behaviour is expected since $\Omega_\Omega h^2$ is inversely proportional to the annihilation cross section which depends on the center-of-mass energy of the colliding non-relativistic DM particles, i.e. $s \approx 4m_\Omega^2$. Here, we are focusing on the case of small doublet-singlet mixing $\theta$, i.e. $\lesssim 0.1$. For this reason, the DM relic abundance turns out to be "blind" to this parameter and is completely determined by the interaction of fields belonging to the dark sector (namely, $\chi$ and/or $h_2$). The DM annihilation channel into a pair of $\chi$ is always present unless further assumptions are made to suppress it.

In contrast to the situation previously described, the DM direct detection given by the SI cross section $\sigma_\Omega$, Eq.~(\ref{dd}), is sensitive to the values of the doublet-singlet mixing $\theta$. This is shown by the plot on the right-hand side of Figure~\ref{figb} where $\sigma_\Omega$ is depicted as a function of the mass of second $\text{CP-even}$ scalar $m_{h_2}$ (considering  $m_\Omega=200$ and 500~GeV, and each case with $\theta = 0.1, 0.01$). The blue and green curves represent the points in the parameter space fulfilling the correct relic abundance while the red, yellow and gray dashed horizontal lines correspond to the experimental limits provided by XENON1T~\cite{XENON:2018voc} and LUX-ZEPLIN (LZ)~\cite{LZ:2022ufs}, and the projections by XENONnT~\cite{XENON:2020kmp}, respectively. 
From Eq.~(\ref{dd}) one can notice that the cross section rests on the existence of a mixing between the scalar doublet $\Phi$ and the singlet $\sigma$. This dependence can be observed from Figure~\ref{figb} which shows the sensitivity of the cross section to $\theta$ variations. Furthermore, we can see that when the CP-even scalars $h_1$ and $h_2$ are (semi)-degenerate, i.e. $m_{h_2} \approx m_{h_1}$, there is a numerical cancellation that generates the inverted peak in the SI cross section. This allows to elude the experimental bounds when $h_1$ and $h_2$ are close in mass. 
 Another possibility to relax the experimental constraints, including the one coming from XENONnT, happens by shrinking the value of doublet-singlet mixing as  depicted in Figure~\ref{figb}. This parameter space limit is reached, for instance, when lepton number gets broken at energies much higher than the electroweak scale. \\

Let us note that this model is characterized by the presence of the process $\Omega\bar{\Omega}\rightarrow \chi h_{1,2}$, with $h_{1,2}$ decaying into SM particles~\cite{Garcia-Cely:2013wda}. These s-wave processes are velocity independent channels not present in other models~\cite{Ahriche:2016acx,Mandal:2019oth} and give good prospects for indirect DM detection especially when $\chi$ is a pseudo-Goldstone boson, i.e. $m_\chi\neq0$. For this reason, we look into regions of the parameter space where the processes $\Omega\bar{\Omega}\rightarrow \chi h_{1,2}$ dominate the DM annihilation and provide the limits coming from the Alpha Magnetic Spectrometer
(AMS) experiment as well as the future sensitivities of the Cherenkov Telescope Array (CTA) experiment. For convenience, we assume $m_\chi>m_{h_2}/2$ so that the decay $h_2 \to 2\chi$ is kinematically disallowed. In this way, the DM candidate does not annihilate primarily into invisible channels. \\

Using Eq.~(\ref{eq:srot}), one can express the $h_2$ branching fraction into SM particles as~\cite{Robens:2015gla}, 
\begin{eqnarray}\label{breq}
 \text{BR} (h_2\rightarrow \text{SM}) = \sin^2\theta \left[\frac{\Gamma (h_2\rightarrow \text{SM})}{\Gamma_{\text{tot}}}\right]
\end{eqnarray}
where $\Gamma (h_2\rightarrow \text{SM})$ corresponds to the partial decay width of the scalar boson $h_2$ (with mass $m_{h_2}$) into SM states, and the total decay width is given by 
\begin{eqnarray}
 \Gamma_\text{tot} = \sin^2\theta \times \Gamma^{\text{SM}}_{\text{tot}} + \Gamma (h_2\rightarrow 2h_1) + \Gamma (h_2\rightarrow 2\Omega) + \Gamma (h_2\rightarrow 2\chi).
 \label{Gtot}
\end{eqnarray}
Here $\Gamma^\text{SM}_{\text{tot}}$ corresponds to the total decay width of $h_2$ into SM states~\cite{LHCHiggsCrossSectionWorkingGroup:2011wcg}. 
For simplicity, we focus on the case in which $m_{h_2} < 2m_\Omega$. Furthermore, in order to assure observability, via $h_2$ decaying into SM particles, we assume $m_\chi\neq0$ as well as $m_{h_2} < 2m_\chi$. Therefore, the third and fourth terms in Eq.~(\ref{Gtot}) are not present in our study.\\

\begin{figure}[t!]
\centering
\includegraphics[width=0.8\textwidth]{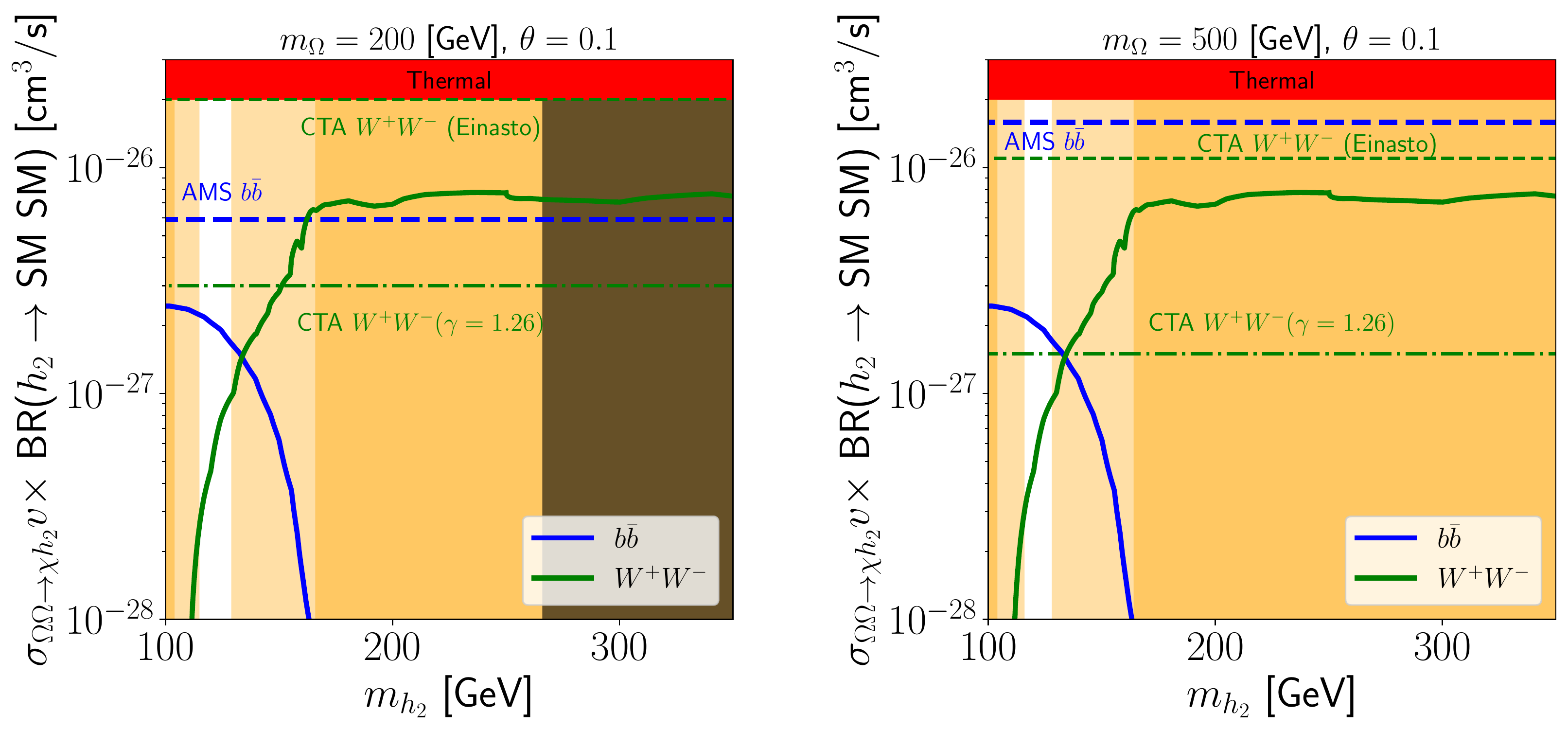} \\
\includegraphics[width=0.8\textwidth]{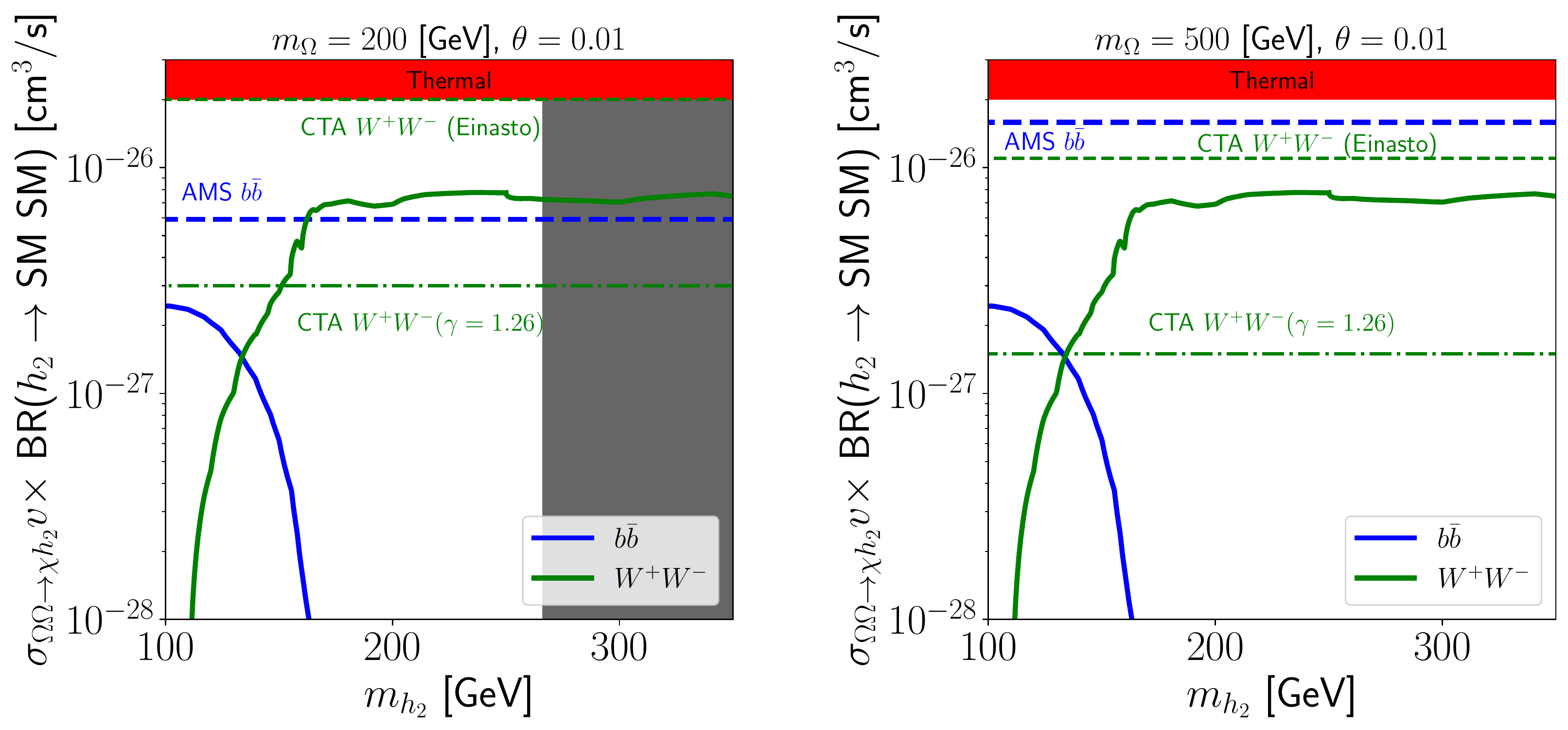}
\caption{Parameter space restrictions from direct and indirect dark matter. The thermal value of $\braket{\sigma v}_{\Omega\Omega\rightarrow h_2\chi}$ is given by the red band on top of the plots. It is assumed that $h_2$ decays into either $b\bar{b}$ (blue solid line) or $W^+W^-$ (green solid line). The dashed blue horizontal line represents the latest bound on the DM annihilation cross section into $b\bar{b}$ from AMS-02. The green dashed and dotted-dashed are the projected sensitivities of CTA depending on the DM profile. The dark and light orange areas correspond to the LZ exclusion region and the future sensitivity of XENONnT, respectively. The dark gray area in the left panels satisfies $m_{h_2} > 2 m_\chi$ and is then forbidden.}
\label{id}
\end{figure}

Taking into account the considerations previously stated, we perform a numerical analysis. Figure~\ref{id} shows the predictions for indirect detection signals coming from the DM annihilation into a pseudo-Goldstone $\chi$ and $h_2$. The CP-even scalar subsequently decays into SM particles, see Eq.~(\ref{breq}). Then, the DM annihilation would be $\Omega\bar{\Omega}\rightarrow \chi h_{2}\rightarrow \chi \text{SM}$. The assumed thermal value $2\times 10^{-26}$ cm$^3/$s$\leq\braket{\sigma v}_{\Omega\bar{\Omega}\rightarrow \chi h_2}\leq 3 \times 10^{-26}$ cm$^3/$s is depicted by the red region at the top of each panel in Figure.~\ref{id}. The left-panels consider $m_\Omega = 200$ GeV for $\theta = 0.1$ and $0.01$, while the panels on the right take $m_\Omega = 500$ GeV for the same values of $\theta$. In all panels, the Yukawa coupling is $y_N = 0.1$, and the dark scalar and pseudoscalar masses are fixed as $m_{\eta_R} = 2000$ GeV and $m_{\eta_I} = 2001$ GeV, respectively. The solid blue (green) line corresponds to the DM matter annihilation into $b\bar{b}$ ($W^+ W^-$). The dashed blue horizontal line is the upper bound of AMS-02~\cite{Reinert:2017aga} for DM annihilation into $b\bar{b}$, whereas the dashed green horizontal one represents the future sensitivity of CTA~\cite{CTA:2020qlo} in the $W^+W^-$ channel. We also consider the bound projected by CTA for DM annihilation searches with $W^+W^-$ in the final state assuming a gNFW profile with a slope parameter $\gamma=1.26$~\cite{CTA:2020qlo}. Figure~\ref{id} includes direct detection bounds at fixed $\theta$ and DM mass. The dark orange is the exclusion region coming from the LZ results on DD searches. The light orange area represents the future sensitivity of XENONnT. Therefore, only the light orange and white parts in each panel are allowed by current DM direct detection constraints. In addition, the dark gray area in the left panels of Figure~\ref{id} satisfies $m_{h_2} > 2 m_\chi$ and is then forbidden. This is because we are working under the assumption that $m_{h_2} < 2 m_\chi$ or $m_\chi=2 m_\Omega - m_{h_2}> 2/3 m_\Omega$, i.e. $h_2$ does not decay into $2\chi$. This guarantees that the fermion DM candidate $\Omega$ annihilates into observable modes.\\

 Figure~\ref{id} shows the parameter space fraction that CTA would be able to test in the future by looking for $W^+W^-$ products. One can see that the CTA projections do not reach the model predictions if the DM density distribution follows an Einasto profile. On the other hand, if the DM posses a gNFW profile, CTA searches could prove masses $m_{h_2}\gtrsim 150$ GeV. In addition, we have that AMS-02 data impose parameter space restrictions over the parameter space from DM annihilation into $b\bar{b}$ searches. Notice that, as expected from Eq.~(\ref{dd}), the direct detection constraints get weaker when the singlet-doublet mixing takes smaller values. 

\section{Charged Lepton Flavor Violation}
\label{clfvandlepto} 

In this section we analyze charged lepton flavor violation (cLFV) processes present due to the mixing between active and heavy sterile neutrinos. Here we focus in the one-loop decays $l_{i}\rightarrow
l_{j}\gamma$ whose branching ratios are given by~\cite{Langacker:1988up,Lavoura:2003xp,Hue:2017lak}
\begin{eqnarray}
\text{BR}\left( l_{i}\rightarrow l_{j}\gamma \right)  
&=&\frac{\alpha
_{W}^{3}s_{W}^{2}m_{l_{i}}^{5}}{256\pi ^{2}m_{W}^{4}\Gamma _{i}}\left\vert
G_{ij}\right\vert ^{2}\label{Brmutoegamma1} \\
G_{ij} &\simeq &\sum_{k=1}^{3}\left( \left[
\left( 1-RR^{\dagger }\right) U_{\nu }\right] ^{\ast }\right) _{ik}\left(
\left( 1-RR^{\dagger }\right) U_{\nu }\right) _{jk}G_{\gamma }\left( \frac{%
m_{\nu _{k}}^{2}}{m_{W}^{2}}\right) +2\sum_{l=1}^{2}\left(R^{\ast }\right)
_{il}\left( R\right) _{jl}G_{\gamma }\left( \frac{m_{N_{R_l}}^{2}}{m_{W}^{2}}%
\right), \label{Brmutoegamma2}\\
G_{\gamma } (x) &=&\frac{10-43x+78x^{2}-49x^{3}+18x^{3}\ln x+4x^{4}}{%
12\left( 1-x\right) ^{4}},  \notag
\end{eqnarray}
where $\Gamma _{\mu }=3\times 10^{-19}$ GeV is the total muon decay width, $U_{\nu}$ is the matrix that diagonalizes the light neutrinos mass matrix which, in our case, is equal to the Pontecorvo–Maki–Nakagawa–Sakata (PMNS) matrix since the charged lepton mixing matrix is equal to the identity $U_{\ell}=\mathbb{I}$. In addition, the matrix $R$ is given by
\begin{equation}
R=\frac{1}{\sqrt{2}}m_D^{*}M^{-1},
\label{eq:Rneutrino}
\end{equation}

 where $M$ and $m_D$ are the heavy Majorana mass matrix and the Dirac neutrino mass matrix, respectively. We provide in Appendix~\ref{apx:mutoegamma} all assumptions made for computing $R$ in Eq.~(\ref{eq:Rneutrino}). Table~\ref{tab:parameter} contains the benchmarks points used to compute $ \mu \rightarrow e\gamma$.\\

Then, feeding Eq.~(\ref{Brmutoegamma1}) with the values of the dimensionless parameters and masses given in Table~\ref{tab:parameter}, one gets the following branching fractions,
\begin{eqnarray}
\text{BR}^{(a)}(\mu \rightarrow e\gamma) \simeq 2.02\times 10^{-13} \ \ \text{and} \ \
\text{BR}^{(b)}(\mu \rightarrow e\gamma) \simeq 1.13\times 10^{-13},
\label{eq:brsbenchmarks}
\end{eqnarray}
where $(a)$ is for $m_{\Omega}=200\text{ GeV}$ and $(b)$ is for $m_{\Omega}=500\text{ GeV}$.\\

\begin{figure}[h!]
\centering
\subfigure[]{
\includegraphics[scale=0.32]{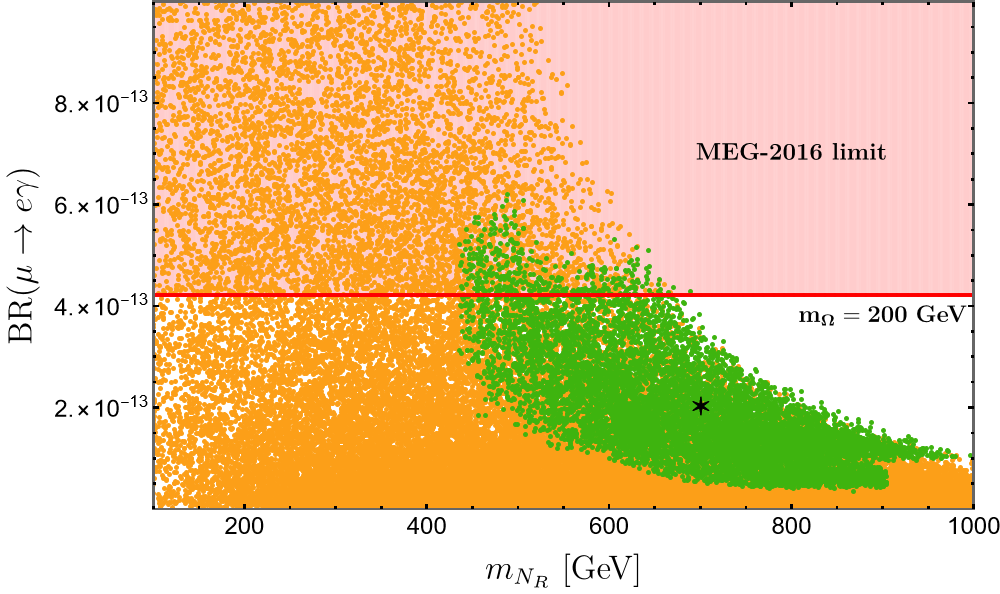}
}
\subfigure[]{
\includegraphics[scale=0.32]{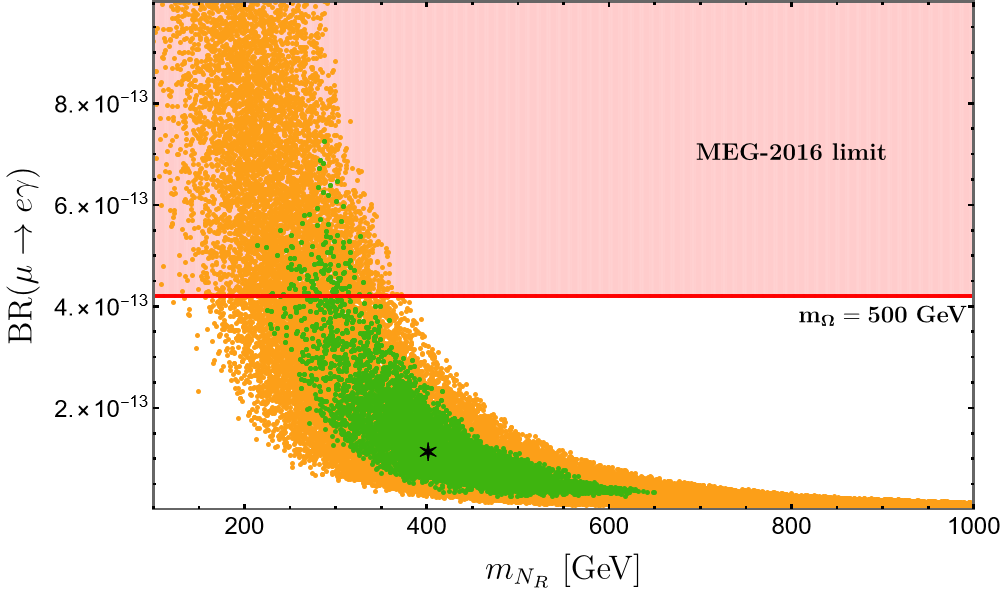}
}
\caption{Branching ratio $\text{BR}\left(\mu \rightarrow e\gamma \right)$ as a function of the mass of the lightest RH Majorana neutrino $N_R$. The shadowed region is excluded excluded by MEG~\cite{MEG:2016leq}. The black star corresponds to the prediction of the best-fit point of the model, for $m_{\Omega}=200\; \text{GeV}$ (left-panel) and $m_{\Omega}=500\; \text{GeV}$ (right-panel). The green points are compatible with current neutrino oscillation experimental limits at $3\sigma$. The orange point are out of the $3\sigma$ range and, hence, excluded by neutrino oscillation data.}

\label{mutoegammas2}
\end{figure}

Figure~\ref{mutoegammas2} shows the correlation between the branching ratio $\text{BR}\left( \mu \rightarrow e\gamma \right)$ and the mass of the lightest RH Majorana neutrino $N_R$. One observes that the branching ratio decreases as the mass of $N_R$ increases. In both plots, the red horizontal line and the shaded region represent the latest experimental constraint provided by the MEG~\cite{MEG:2016leq} collaboration,
\begin{equation}
\text{BR}\left( \mu \rightarrow e\gamma \right)^{\text{exp}}<4.2\times 10^{-13}.
\end{equation}

The black stars in Figure~\ref{mutoegammas2} correspond to the branching ratio predicted, Eq.~(\ref{eq:brsbenchmarks}), by the best-fit points of the model for $m_{\Omega}=200\; \text{GeV}$ (left-panel) and $m_{\Omega}=500\; \text{GeV}$ (right-panel). The scatter plots come from a random variation of the dimensionless parameters up to 30\% around the best-fit value. The green points are compatible with current neutrino oscillation experimental limits at $3\sigma$. One can see that neutrino oscillation data restrict the lightest RH neutrino mass to be in the range $436.9\;\text{GeV}\leq m_{ N_R}\leq 996.5 \; \text{GeV}$ for $m_{\Omega}=200\; \text{GeV}$, and $204.5\;\text{GeV}\leq m_{ N_R}\leq 649.9 \; \text{GeV}$ for $m_{\Omega}=500\; \text{GeV}$. All orange points are out of the $3\sigma$ range and, hence, excluded by neutrino oscillation data. 

\newpage

\section{Conclusions \label{sec:conc}}
We have built an SM extension where the tiny masses of the
light active neutrinos are generated from an inverse seesaw mechanism at one-loop level. This model adds two scalar singlets and six right handed Majorana neutrinos to the SM particle content.
In addition, the SM gauge group is enlarged by 
a global $U(1)_X$ symmetry, which is spontaneously broken down to an Abelian discrete subgroup, $\mathcal{Z}_2\subset U(1)_X$.  
The latter is an exact low-energy symmetry guarantying a one-loop realization of the inverse seesaw mechanism and the existence of stable scalar and fermionic dark matter candidates. We focused our study on the case in which the dark matter is the stable Majorana fermion.\\

We have found that our model, besides providing a dynamical origin of the inverse seesaw mechanism, successfully reproduces the measured values of the DM relic density. We have identified the parameter space regions sensitive to (in)direct DM searches. In the case of direct detection searches, we show the parameter space restrictions of the model (imposed by XENON1T and LZ results) and future prospects of XENONnT experiment. Furthermore, we compare the model predictions coming from the Majorana fermion annihilation into $b\bar{b}$ and $W^+W^-$ with AMS-02 bounds and CTA projections. We found that our benchmarks could be tested by CTA if the DM has a gNWA profile.\\ 

We have also analysed the lepton flavor violating $\mu\to e \gamma$ process and provided the model predictions in simplified scenarios that are in agreement with current neutrino oscillation data.

\section*{Acknowledgments}
The work of C.B. was supported by FONDECYT grant No. 11201240.
A.E.C.H is supported by ANID-Chile FONDECYT 1210378, ANID PIA/APOYO AFB220004 and Milenio-ANID-ICN2019\_044. S.K is supported by ANID-Chile FONDECYT 1230160 and Milenio-ANID-ICN2019\_044. J.M. is supported by ANID Programa de Becas Doctorado Nacional code 21212041. 
B.D.S has been founded by ANID Grant No. 74200120.
C.B. would like to acknowledge
the hospitality and support from the ICTP through the Associates Programme (2023-2028).
\appendix
\section{Boundedness Conditions}
\label{apx:stabilityconds}

The boundedness conditions of the scalar potential in Eq.~(\ref{eq:pot}) are
derived assuming that the quartic terms dominate over at high energies. In order to do so, we define the following bilinears,
\begin{equation}
a=|\Phi |^{2}\quad ;\quad b=|\sigma |^{2}\quad ;\quad c=|\eta |^{2},
\label{eq:bcpar}
\end{equation}
and rewrite the quartic terms of the scalar potential. Then, using the expressions in Eq.~(\ref{eq:bcpar}) one gets
\begin{eqnarray}
V_{q} &=&\frac{1}{2}(\sqrt{\lambda _{\Phi }}a-\sqrt{\lambda _{\sigma }}%
b)^{2}+\frac{1}{2}(\sqrt{\lambda _{\Phi }}a-\sqrt{\lambda _{\eta }}c)^{2}+%
\frac{1}{2}(\sqrt{\lambda _{\sigma }}b-\sqrt{\lambda _{\eta }}%
c)^{2}+(\lambda _{1}+\sqrt{\lambda _{\Phi }\lambda _{\sigma }})ab  \notag \\
&&+(\lambda _{2}+\sqrt{\lambda _{\Phi }\lambda _{\eta }})ac+(\lambda _{3}+%
\sqrt{\lambda _{\sigma }\lambda _{\eta }})bc-\frac{1}{2}(\lambda _{\Phi
}a^{2}+\lambda _{\eta }c^{2}).  \label{ec:potentialcuartic}
\end{eqnarray}

Following  Refs.~\cite{Maniatis:2006fs,Bhattacharyya:2015nca} the boundedness conditions
of the model turn out to be,
\begin{eqnarray}
\lambda _{\Phi } &\geq &0\quad ;\quad \lambda _{\sigma }\geq 0\quad ;\quad
\lambda _{\eta }\geq 0  \label{ec:condition1} \\
\lambda _{1}+\sqrt{\lambda _{\Phi }\lambda _{\sigma }} &\geq &0\quad ;\quad
\lambda _{2}+\sqrt{\lambda _{\Phi }\lambda _{\eta }}\geq 0\quad ;\quad
\lambda _{3}+\sqrt{\lambda _{\sigma }\lambda _{\eta }}\geq 0
\label{ec:condition2}
\end{eqnarray}

\section{Benchmarks for $\mu\to e \gamma$ }
\label{apx:mutoegamma}

The cLFV process $\mu \rightarrow e\gamma$ is computed using Eq.~(\ref{Brmutoegamma2}) and taking as inputs the model outputs that minimize the $\chi^2$ function given by,
\begin{equation}
\chi ^{2}=\frac{\left[ \Delta m_{21}^{2\,(\exp)}-\Delta m_{21}^{2\,(\text{th})}\right] ^{2}}{\sigma
_{\Delta m_{21}^2}^{2}}+\frac{\left[ \Delta m_{31}^{2\,(\exp)}-\Delta m_{31}^{2\,(\text{th})}\right] ^{2}}{\sigma
_{\Delta m_{31}^2}^{2}}+\sum_{i<j}\frac{\left[ s_{{ij}}^{(\exp)}-s_{ij}^{(\text{th})}\right] ^{2}}{\sigma _{s_{ij}}^{2}}+\frac{\left[ \delta _{CP}^{(\exp) }-\delta
_{CP}^{(\text{th})}\right] ^{2}}{\sigma _{\delta_{CP} }^{2}}\;,  \label{eq:chisq}
\end{equation}
where $s_{ij}\equiv \sin\theta_{ij}$ (with $i,j=1,2,3$), $\delta_{CP}$ is the leptonic CP violating phase, the label (th) are used to identify the model outputs, while the ones with label (exp) correspond to the experimental values, and $\sigma_a $ represent the experimental errors. Table~\ref{table:neutrinos_data} shows best fit values and $1\sigma-3\sigma$ intervals reported by neutrino oscillation global fits~\cite{deSalas:2020pgw} \footnote{Table~\ref{table:neutrinos_data} correspond to normal neutrino mass ordering. The inverted mass ordering can be consulted in ~\cite{deSalas:2020pgw}. For other fits of neutrino oscillation parameters we refer the reader to Refs.~\cite{Capozzi:2017ipn, Esteban:2020cvm}.}.

\begin{table}[h!]
\begin{tabular}{c|cccccc} 
\hline\hline
\text{Observable}      & \text{$\Delta m_{21}^2[10^{-5}\;\text{eV}]$} & \text{$\Delta m_{31}^2[10^{-3}\;\text{eV}]$} & \text{$\sin^2\theta_{12}/10^{-1}$} & \text{$\sin^2\theta_{23}/10^{-1}$} & \text{$\sin^2\theta_{13}/10^{-2}$} & \text{$\delta_{\text{CP}}/^{\circ}$}  \\ 
\hline\hline
\text{ Best fit $\pm1\sigma$} & $7.50_{-0.20}^{+0.22}$                         & $2.55_{-0.03}^{+0.02}$                         & $3.18\pm 0.16$                       & $5.74\pm 0.14$                       & $2.200_{-0.062}^{+0.069}$            & $194_{-22}^{+24}$                   \\
\text{$3\sigma$ range} & $6.94-8.14$                                    & $2.47-2.63$                                    & $2.71-3.69$                          & $4.34-6.10$                          & $2.00-2.405$                         & $128-359$                           \\
\hline\hline
\end{tabular}
\caption{Neutrino oscillation parameters from global fits~\cite{deSalas:2020pgw}.}
\label{table:neutrinos_data}
\end{table}

In order to compute all neutrino oscillation parameters we perform a random scan of the free parameters in the lepton sector and make assumptions about the flavor structure of the Dirac mass matrix $m_{D}$, and the Majorana matrices $M$ and $\mu$ in Eq.~(\ref{Mnu}). Using the Casas–Ibarra parametrization~\cite{Casas:2001sr}, the matrix $m_{D}$ in Eq.~\eqref{ec:mdirac}, reads as follows~\cite{Casas:2001sr,Ibarra:2003up,Restrepo:2019ilz,Cordero-Carrion:2019qtu,Dolan:2018qpy,Hernandez:2021xet},
\begin{equation}
m_{D}=  \frac{y_\nu v_\Phi}{\sqrt{2}}=U_{\text{PMNS}}\left(\hat{m}_{\nu}\right)^{1/2}\hat{R}\mu^{-1/2}M\;,  \label{eq:mD-casa}
\end{equation}%
where $U_{\text{PMNS}}\equiv U_\ell^{\dagger}U_\nu$, $\hat{m}_{\nu}=\text{diag}(m_1,m_2,m_3)$ is the diagonal neutrino mass matrix and $\hat{R}$ is a rotation matrix given by,
\begin{eqnarray}
\hat{R}=\left(
\begin{tabular}{cc}
0 & 0 \\
$\cos\hat{\theta}$ & $\sin\hat{\theta}$ \\
$-\sin\hat{\theta}$ & $\cos\hat{\theta}$
\end{tabular}
\right)\ \ \text{with} \ \ \hat{\theta}\in[0,2\pi]. \label{ec:Rot}
\end{eqnarray}

The matrices $M$ and $\mu $ are assumed to be diagonal,
\begin{eqnarray}
M=\left(
\begin{tabular}{cc}
$M_1$ & 0 \\
0 & $M_2$
\end{tabular}
\right)\ \ \text{and} \ \
\mu=\left(
\begin{tabular}{cc}
$\mu_1$ & 0 \\
0 & $\mu_2$
\end{tabular}
\right)
.\label{ec:benchmark}
\end{eqnarray}

For the matrix $M$ in Eq.~(\ref{ec:benchmark}) we varied $M_1$ within the range $100\;\text{GeV} \leq M_{1} \leq 1\; \text{TeV}$ and considered $M_2= 10 M_1$, where $M_1$ is the mass of the lightest RH neutrino, $M_1=m_{N_R}$. In the case of the $\mu$ matrix we used Eq.~\eqref{eq:matrix-mu} to compute $\mu_1$ and assumed $\mu_2=10\mu_1$. This matrix depends on $y_N$, $m_{\Omega}$, $m_{\eta_{R}}$ and $m_{\eta_{I}}$. Then, following the study made in section~\ref{DM}, we analyze two situations: one with DM mass of $m_{\Omega}=200$ and the other with $m_{\Omega}=500\; \text{GeV}$. In both cases, the Yukawa $y_N$ was varied in the range $10^{-2}\leq y_N \leq 1$ and the masses of the $\mathcal{Z}_2$-odd scalars were fixed to
\begin{eqnarray}
 m_{\eta_R}=2000\; \text{GeV}, \ \ m_{\eta_I}=2001\; \text{GeV}.  
 \end{eqnarray}

We also take the charged lepton mass matrix as a diagonal matrix, i.e. $M_{l}=\text{diag}(m_e,m_\mu,m_\tau)$, which implies $U_{\text{PMNS}}=U_\nu$ in Eq.~(\ref{eq:mD-casa}).\\

 Table~\ref{tab:parameter} shows the dimensionless parameters that minimize the $\chi^2$ function in Eq.~(\ref{eq:chisq}) and that are used to compute $\mu\to e\gamma$, Eq.~(\ref{Brmutoegamma2}). The best fits of the model, for $m_{\Omega}=200\text{ GeV}$ and $m_{\Omega}=500\text{ GeV}$ are presented in Table~\ref{table:modelfit}.\\

\begin{table}[h!]
\centering
\begin{tabular}{l|l} 
\hline\hline
\multicolumn{2}{c}{\textbf{Dimensionless parameters $(a)$}}                \\ 
\hline\hline
$y_{\nu_{11}}=0.0109e^{1.87i}$  & $y_{\nu_{31}}=0.0124e^{1.49i}$  \\
$y_{\nu_{12}}=0.0105e^{-1.31i}$ & $y_{\nu_{32}}=0.0808e^{1.77i}$  \\
$y_{\nu_{21}}=0.0270e^{1.70i}$  & $y_N=2.04\times 10^{-2}$        \\
$y_{\nu_{22}}=0.0463e^{1.56i}$  & $\hat{\theta}=1.303\; \text{rad}$     \\ 
\hline\hline
\end{tabular}
\qquad
\begin{tabular}{l|l} 
\hline\hline
\multicolumn{2}{c}{\textbf{Dimensionless parameters $(b)$}}                  \\ 
\hline\hline
$y_{\nu_{11}}=0.00525e^{-1.51i}$ & $y_{\nu_{31}}=0.0177e^{1.572i}$  \\
$y_{\nu_{12}}=0.0151e^{-1.59i}$  & $y_{\nu_{32}}=0.00424e^{1.61i}$  \\
$y_{\nu_{21}}=0.0166e^{1.57i}$   & $y_N=1.22\times 10^{-2}$         \\
$y_{\nu_{22}}=0.0354e^{-1.58i}$  & $\hat{\theta}=-4.98\; \text{rad}$      \\ 
\hline\hline
\end{tabular}
\caption{Dimensionless parameters used to compute $\text{BR}(\mu\rightarrow e\gamma)$ compatible with neutrino oscillation data. Case (a) considers $m_{\Omega}=200$  GeV and case (b) is for $m_{\Omega}=500$  GeV.}
\label{tab:parameter}
\end{table}

\begin{table}[h!]
\begin{tabular}{c|cccccccc}
\hline\hline
Observable & $\mu_1[\text{keV}]$ & $m_{N_R}[\text{GeV}]$ & $\Delta m_{21}^{2}$[$10^{-5}$eV$^{2}$]
& $\Delta m_{31}^{2}$[$10^{-3}$eV$^{2}$] & $\sin\theta^{(l)}_{12}/10^{-1}$
& $\sin\theta^{(l)}_{13}/10^{-3}$ & $\sin\theta^{(l)}_{23}/10^{-1}$ & $%
\delta^{(l)}_{CP} (^{\circ })$ \\ \hline\hline
Best fit case (a)  &  $-0.562$ & $700.5$ & $7.53$ & $2.51$ & $3.53$ & $2.11$ & $5.66$ & $195.3$\\ \hline
Best fit case (b) & $-0.4099$ & $401.3$ & $7.50$ & $2.55$ & $3.25$ & $2.22$ & $5.64$ & $174.8$\\ \hline\hline
\end{tabular}
\caption{Best fit values of the model. Case (a) considers $m_{\Omega}=200\text{ GeV}$ and case (b) is for $m_{\Omega}=500\text{ GeV}$.}
\label{table:modelfit}
\end{table}

\section*{References}
\bibliographystyle{utphys}
\bibliography{Refs.bib}

\end{document}

%% file: diagram.tex
\usepackage{graphicx}
\usepackage{bm}
\usepackage{float}
\usepackage{subfigure}
\usepackage{tikz}
\usetikzlibrary{arrows.meta}
\usetikzlibrary{shapes.geometric}

\usetikzlibrary{arrows,shapes}
\usetikzlibrary{trees}
\usetikzlibrary{matrix,arrows} 				
\usetikzlibrary{positioning}				
\usetikzlibrary{calc,through}				
\usetikzlibrary{decorations.pathreplacing}  
\usepackage{pgffor}							

\usetikzlibrary{decorations.pathmorphing}	
\usetikzlibrary{decorations.markings}

\tikzset{
    vector/.style={decorate, decoration={snake}, draw},
	provector/.style={decorate, decoration={snake,amplitude=2.5pt}, draw},
	antivector/.style={decorate, decoration={snake,amplitude=-2.5pt}, draw},
    fermion/.style={draw=black, postaction={decorate},
        decoration={markings,mark=at position .55 with {\arrow[draw=black]{>}}}},
    fermionbar/.style={draw=black, postaction={decorate},
        decoration={markings,mark=at position .55 with {\arrow[draw=black]{<}}}},
    fermionnoarrow/.style={draw=black},
    gluon/.style={decorate, draw=black,
        decoration={coil,amplitude=4pt, segment length=5pt}},
    scalar/.style={dashed,draw=black, postaction={decorate},
        decoration={markings,mark=at position .55 with {\arrow[draw=black]{>}}}},
    scalarbar/.style={dashed,draw=black, postaction={decorate},
        decoration={markings,mark=at position .55 with {\arrow[draw=black]{<}}}},
    scalarnoarrow/.style={dashed,draw=black},
    electron/.style={draw=black, postaction={decorate},
        decoration={markings,mark=at position .55 with {\arrow[draw=black]{>}}}},
	bigvector/.style={decorate, decoration={snake,amplitude=4pt}, draw},
}
\usetikzlibrary{decorations.markings}